# The activated torsion oscillation magnetometer

G. Asti, M. Ghidini[1], R. Pellicelli, M. Solzi

*INFM-Dipartimento di Fisica dell'Università di Parma, Parco Area delle Scienze 7/A, I-43100 Parma, Italy*

**Abstract**

*The activated torsion oscillation magnetometer exploits the mechanical resonance of a cantilever beam, driven by the torque exerted on the sample by an ac field applied perpendicularly to the film plane. We describe a model for the cantilever dynamics which leads to the calculation of the cantilever dynamic profile and allows the mechanical sensitivity of the instrument to be expressed in terms of the minimum electronically detectable displacement. We have developed a capacitance detector which is able to detect displacements of the order of 0.1 nm. We show that sensitivities of the order of $0.5 \times 10^{-11}$ $Am^2$ can be in principle achieved. We will subsequently describe the main features of the ATOM prototype which we have built and tested, with particular attention to the design solutions which have been adopted in order to reduce the effects of parasitic vibrations due either to acoustic noise or to eddy currents. The instrument is mounted in a continuous flow cryostat and can work in the 4.2-300 K temperature range. Finally, we will show that our experimental set-up has a second mode of operation, named Torsion Induction Magnetometer (TIM).*

*Keywords*: Torque Magnetometry, Magnetometry, Anisotropy-thin films, Thin-films-magnetic

## 1. Introduction

The present emphasis towards artificial nanostructured magnetic materials calls for suitable measuring techniques especially in terms of enhanced sensitivity. For the measurement of macroscopic magnetic properties in oligatomic samples, SQUID and MOKE magnetometers are widely employed, as well as VSM and AGFM, with some limitations. Torque magnetometry is a classical technique of inherent high sensitivity for measuring both the magnetic moment and the first order anisotropy constant [1] which has recently actracted interest for application to magnetic thin films. In the case of thin films, torsion magnetometers exploit the torque exerted on the sample in a homogeneous magnetic field due to its magnetic anisotropy, mainly of shape in origin. The dia- or paramagnetic contribution of the substrate is strongly reduced (see further Sect.II) both with respect magnetic force magnetometers (e.g. AGFM) and induction magnetometers (e.g. VSM, SQUID).

Torque magnetometers for thin films have been implemented either in the static mode (TM) [2] or, as torsion pendulums, in the oscillating mode (TOM) [3]. In the case of the TM the magnetic moment can be deduced by directly measuring the torque exerted on the sample as a function of the applied static magnetic field. In the case of the TOM the magnetic moment can be deduced from the magnetic stiffness, which is determined by measuring the frequency shift of the free or forced oscillation of the pendulum as a function of the magnetizing field. The oscillation is not activated through the sample magnetic moment. Gradmann et al. demonstrated the high sensitivity of TOM and provided a detailed analysis of the two methods and of the attainable sensitivities [3]. Recently, torque magnetometry has actracted renewed interest in connection with the development of advanced micromachining techniques. Silicon cantilever beams and rods in the micrometric and submicrometric size ranges can be fabricated and utilized as highly sensitive torsion elements. Furthermore, micromachined cantilevers can be easily combined with different sensors (e.g. optical, capacitive, piezoresistive) and actuators. Both TM and TOM have been developed employing such torsion elements attaining high sensitivities. In the current literature one can distinguish between macroscopic cantilevers which are able to host a thin film deposited onto a given substrate [4-7], and micro-cantilever beams onto which samples can be directly deposited by Molecular Beam Epitaxy or other thin film deposition techniques[8-12]. However, only a few examples of magnetically activated torsion pendulum have been reported so far in the literature [12,13].

In the following, the principles and the characteristics of an Activated Torsion Oscillation Magnetometer (ATOM) are analyzed. The working principle is that of a torsion system (a macroscopic cantilever beam) driven in resonance by the torque exerted on a magnetic thin film by a small ac field applied perpendicularly to the film plane and to a dc magnetizing field. A detailed analysis shows that such instrument can achieve very high sensitivity. An ATOM prototype is described and tested. The prototype is able to host any substrate/sample system and therefore lends itself to general application as an ex-situ instrument.

---

[1] Corresponding author. Tel.: +39-0521-905276; fax: +-39-0521-905223; e-mail: ghidini@fis.unipr.it

## II. Basic principles

Let us consider the sample as a physical system exchanging energy with the environment, under the action of an external magnetic field $\vec{H}_e$. Then we have from the first principle of thermodynamics

$$-\int \vec{H}_e \cdot d\vec{I} dv + \delta W_m + dU = \delta Q \quad (1)$$

where $\vec{I} = \mu_0 \vec{M}$ represents the magnetic polarization, $\delta W_m$ is the elementary mechanical work, $U$ is the sample internal energy including the magnetic dipolar energy $U_s$ of its magnetic field $H_s$ (demagnetizing field) and $\delta Q$ is the heat absorbed by the system. It is also possible to consider as a physical system the sample without the field $H_s$ so as to eliminate the dependence on the sample shape. This means that we consider only the intrinsic energy of the sample $U'=U-U_s$, thus excluding the dipolar energy.

$$-\int \vec{H}_e \cdot d\vec{I} dv = -\int \vec{H} \cdot d\vec{I} dv + \int \vec{H}_s \cdot d\vec{I} dv = -\int \vec{H} \cdot d\vec{I} dv - dU_s \quad (2)$$

Then with these substitutions eq. (1) becomes

$$-\int \vec{H} \cdot d\vec{I} dv + \delta W_m + dU' = \delta Q \quad (3)$$

If we are interested in calculating the force and torque on a magnetic body under the action of a constant external field it is convenient to start from eq. (1). The free enthalpy is

$$G = F - \int \vec{H}_e \cdot \vec{I} dv = U - TS - \int \vec{H}_e \cdot \vec{I} dv, \quad (4)$$

where F is the free energy, U is the internal energy, T is the temperature and S is the entropy. For a reversible transformation at constant temperature we have

$$dG + \delta W_m + \int \vec{I} \cdot d\vec{H}_e dv = 0 \quad (5)$$

Hence the expressions of the torque $\vec{\tau}$ and the force $\vec{F}$ at constant field $\vec{H}_e$ turn out to be

$$\vec{\tau} = -\frac{\partial G}{\partial \vec{\theta}} \quad ; \quad \vec{F} = -\frac{\partial G}{\partial \vec{r}} \quad (6)$$

Let us estimate the torque on a typical substrate having the shape of a disk. We approximate it with an oblate ellipsoid of revolution having semiaxes a=b and c parallel to the axes x,y and z respectively. We suppose it to be free to rotate around axis b and that field $\vec{H}_e$ lies in the xz plane at an angle θ with respect to axis a. Then $\vec{I}$ also lies in the xz plane and forms with the a-axis an angle γ. Then

$$\frac{G}{V} = -\vec{H}_e \cdot \vec{I} + \frac{I^2}{2\mu_0}(N_a \cos^2 \gamma + N_c \sin^2 \gamma) + \frac{\mu_0}{2}\chi H_e^2 \quad (7)$$

where $N_a$ and $N_c$ are the demagnetizing factors along a and c axes.

Taking into account that $\vec{I} = \mu_0 \chi \vec{H}$ and $\vec{H} = \vec{H}_e - (\overline{\overline{N}}/\mu_0)\vec{I}$ we can express G in terms of $\vec{H}_e$:

$$\frac{G}{V} = -\frac{\mu_0}{2}H^2_e\left(\frac{\chi}{1+\chi N_a}\cos^2\theta + \frac{\chi}{1+\chi N_c}\sin^2\theta\right) \quad (8)$$

Hence the torque per unit volume tuns out to be

$$\tau = \mu_0 \chi^2 H^2_e \frac{N_a - N_c}{(1+\chi N_a)(1+\chi N_c)}\sin\theta\cos\theta \quad (9)$$

Note that the torque for a material of constant susceptibility is proportional to the square of the susceptibility. As a consequence in the case of a dia- or para-magnetic substrate the background torque $\tau = \tau_s V$ is negligible. Let us make a direct comparison with a typical torque $\tau_f$ of a ferromagnetic thin film sample having magnetic dipole moment μ. The ratio of the maximum torques turns out to be

$$\frac{\tau_s}{\tau_f} = \frac{V\chi^2 H_x (N_{xx} - N_{zz})}{\mu(1+\chi N_{zz})(1+\chi N_{xx})} \quad (10)$$

If χ is << 1 , and if the substrate is thin we can approximate the ratio as follows

$$\frac{\tau_s}{\tau_f} \cong \frac{H_x}{M} \cdot \frac{c}{d} \cdot \chi^2 \quad (11)$$

where d is the thickness of the film and M is the sample magnetization. Let us consider the following values d=1 nm, c=0.1 mm, $\chi^2 = 10^{-9}$, M=1 MA/m, $H_x$= 1 MA/m. We obtain $\tau_s/\tau_f = 10^{-4}$. Note that we have considered a very hard ferromagnetic film. With soft materials the situation is even more favorable.

## III. Cantilever dynamics

In this section we assess the problem of the dynamics of the cantilever in the ATOM. We consider a beam of length L, width a, thickness b and Young's modulus E. A magnetic sample is attached at the free end of the beam and is submitted to the combined action of both a static (dc) and a time-dependent (ac) magnetic field, which we write as

$\vec{H}_x = H_0 \vec{i}$ and $\vec{H}_y = H_y \vec{j}$ respectively. We analyze the problem assuming small displacements from equilibrium and that the mass of the cantilever is negligible with respect to the mass M of the sample, which is supposed to have zero moment of inertia. Because we assume the cantilever to be very thin, we also neglect shear deformation. If the sample has a magnetic dipole moment $\vec{\mu} = \mu_0 \vec{m}$, it will experience a torque:

$$\vec{\Gamma}_z = (\mu_x H_y - \mu_y H_0)\vec{k} = (\mu_x H_y - \mu_y H_0)\vec{k} \quad (13)$$

if we consider that the cantilever is sensitive only to the z component of the total torque. The sample fixed on the free end of the beam exerts on it both an inertial force and a torque during motion. The equilibrium conditions, for a portion of the cantilever beam comprised between the free end and a generic point located at a distance x from the clamped end (Fig.1), are

$$\vec{R} + \vec{F} = 0; \quad \vec{T} + (L-x)\vec{i} \times \vec{F} + \vec{\Gamma} = 0 \quad (14)$$

where

$$\vec{F} = F_y \vec{j} = (-m\ddot{y}_L - \beta \dot{y}_L)\vec{j} \quad (15)$$

is the force applied by the sample on the free end (x=L) of the beam, and $\vec{R}$ and $\vec{T}$ are the resistant force and the torque exerted by the constraint at the point x respectively. The curvature of the beam in the point x turns out to be:

$$y'' \cong \frac{1}{\rho} = -\alpha T_z = -\frac{1}{EI} T_z \quad (17)$$

where $I = ab^3/12$ is the second moment of the rectangular cross-section of the beam. Taking into account Eq.(14), one obtains:

$$y'' = \alpha\{(L-x)F_y + \Gamma\} \quad (18)$$

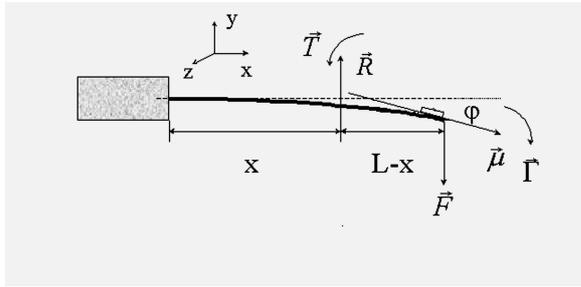

**Fig.1** Schematic representation of the bending of the cantilever beam, onto which a sample of magnetic moment $\vec{\mu}$ and concentrated mass M is attached at the free end (x=L), when submitted to both an active torque $\vec{\Gamma}$ and an active force $\vec{F}$. $\vec{R}$ and $\vec{T}$ are respectively the resistant force and torque exerted by the constraint at the point x.

By double integrating the above equation with respect to x, one gets

$$y = \left(\alpha L \frac{x^2}{2} - \frac{\alpha}{6}x^3\right)F_y + \alpha \Gamma_z \frac{x^2}{2} \quad (19)$$

It should be noted that there is a significant difference with respect to the static case. In fact in the absence of the inertial force, the curve would be a simple arc of a circle. We now focus our attention on the free end of the beam, x=L, and we derive its equation of motion. Being the rotation angle φ of the free end of the beam very small, we can assume $\mu_y \approx \mu\varphi$, $\mu_x \approx \mu$ and y'(x=L)= y'$_L$≈φ. So Eq. (13) can be written as

$$\tilde{\Gamma}_z \cong -\mu y'_L H_0 + \mu H_1 e^{i\omega t} = -\frac{\mu H_0}{EI}\left(\frac{\tilde{F}_y L^2}{2} + \tilde{\Gamma}_z L\right) + \mu H_1 e^{i\omega t} \quad (20)$$

where we have considered an ac field $H_y = H_1 e^{i\omega t}$. From the above, simple algebra yields $\tilde{\Gamma}_z$ as a function of $\tilde{F}_y$,

$$\tilde{\Gamma}_z = \frac{\mu H_1 e^{i\omega t}}{1+\varepsilon} - \frac{L\varepsilon \tilde{F}_y}{2(1+\varepsilon)} \quad (21)$$

where $\varepsilon = \mu H_0 L/EI$. In Eq.(19) we now substitute x=L:

$$y_L = \frac{l^3}{3EI}F_y + \frac{l^2}{2EI}\Gamma_z \quad (22)$$

Finally, substituting Eq. (15) and (21), one obtains the equation of motion:

$$M\frac{d^2\tilde{y}_L}{dt^2} + \beta\frac{d\tilde{y}_L}{dt} + K\tilde{y}_L = Ae^{i\omega t} \quad (23)$$

where $K = 12EI(1+\varepsilon)/[L^3(4+\varepsilon)]$, $A = 6\mu H_1/[L(4+\varepsilon)]$.
The above analysis allows the dynamic profile of the beam to be calculated as a function of x at different instants of time. In fact, at resonance ($\omega = \omega_0$):

$$\tilde{F}_y = -M\ddot{\tilde{y}}_L - \beta\dot{\tilde{y}}_L = -Ae^{i\omega_0 t}\left(1 + i\frac{\omega_0 M}{\beta}\right) = -Ae^{i\omega_0 t}(1+iQ) \quad (24)$$

where Q is the quality factor. An expression for the profile $\tilde{y}(x)$ in which the dependence on time is explicit can be finally obtained. Upon substitution of (24) and (21) in (19) we have:

$$\tilde{y} = -Ae^{i\omega_0 t}(1+iQ)\left[\frac{\alpha L x^2(2+\varepsilon)}{4(1+\varepsilon)} - \frac{\alpha}{6}x^3\right] + \frac{\alpha x^2 \mu_x H_1 e^{j\omega_0 t}}{2+\varepsilon} \quad (25)$$

Figure 2 reports the imaginary part of (25) at different instants of time. The calculation is carried out for a cantilever carrying a sample of mass M=$10^{-6}$ Kg with a magnetic moment of $5.10^{-10}$ A.m$^2$. and H$_1$=1 kA/m. We assumed the beam to display a resonance with Q =100 and an angular frequency of 3164 rad/s. It can be observed that dynamic effects, resulting in an S-shaped profile, are most evident when the beam is swinging through the x axis, i.e. when y$_L$ is small. The above model of cantilever dynamics allows the mechanical sensitivity of the ATOM to be quantitatively evaluated. In fact, solving (23) at resonance, yields the following expression for the displacement amplitude:

$$Y_L = |\tilde{y}_L| \cong \frac{6Q\mu H_1}{KL(4+\varepsilon)} \quad . \quad (26)$$

which yields the expression of the sensitivity in the limit of small magnetic moments (m$_x \to 0$, and hence $\varepsilon \to 0$):

$$S_A = \left|\frac{\partial Y_L}{\partial \mu_x}\right| = \frac{3QH_1}{2LK} \quad . \quad (27)$$

Substituting plausible values of the parameters, i.e. K≈10 N/m, H$_1$=1 kA/m, L = 10 mm and Q=$10^3$ , the above expression yields a minimum detectable magnetic moment $\delta m_x$=5.3×$10^{-2}$ $\delta Y_L$. In principle, supposing to be able to detect oscillations $\delta Y_L$ of the order of 0.1 nm (see further, Sect. IV), one gets a moment resolution of 0.5×$10^{-11}$ Am$^2$, which is the equivalent of the saturation moment of an iron monolayer of area 0.04 mm$^2$. The sensitivity of the ATOM can be compared with the sensitivity of the TOM once we

$$Y_L \approx \frac{F}{M\omega_0\sqrt{4(\omega-\omega_0)^2 + \omega_1^2}} \quad (28)$$

where $\omega_1=\beta/M$. The TOM sensitivity can be evaluated as:

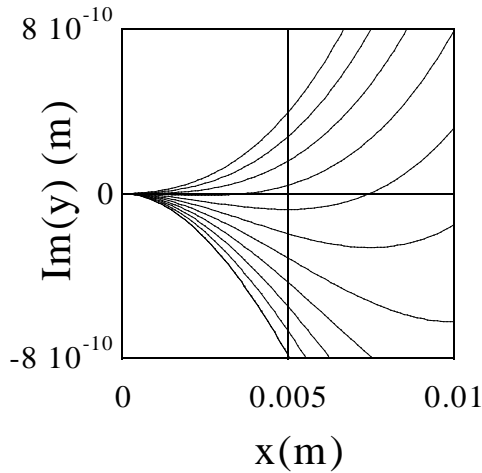

**Fig.2** Detailed view (for small vibration amplitudes) of the imaginary part of the cantilever profile (Eq. (25)), at different instants of time.

realize that the same cantilever can be utilized as the torsion element of a TOM. Near resonance, the modulus of the displacement amplitude can be approximated as:

$$S_T = \left|\frac{\partial Y_L}{\partial \omega_0}\right|_{\omega=\omega^*} \cdot \left.\frac{\partial \omega_0}{\partial \mu_x}\right|_{\mu_x \to 0} \approx \frac{3\sqrt{2}Y_{L0}H_0 Q}{4KL^2} \quad (29)$$

where $\omega_0$ is the resonance frequency of the cantilever and $\omega^* = \omega_0\left\{1 \pm \left(2\sqrt{2}Q\right)^{-1}\right\}$, are the inflexion points of the resonance curve (where the amplitude variation corresponding to a given frequency shift is maximum) and Y$_{L0}$ is the maximum allowed displacement for which the whole system preserves linearity. The ratio of the ATOM to TOM sensitivity gives

$$\frac{S_A}{S_T} = \frac{\sqrt{2}LH_1}{Y_{L0}H_0} \quad (30)$$

Linearity requires that the ratio Y$_{L0}$/L be much smaller than 1 (a typical value inferred from practice is $10^{-3}$). H$_1$ is only limited by the requirement that the longitudinal component of the magnetic moment undergoes small reversible changes. On the other hand H$_0$ can be either smaller than H$_1$ or several orders of magnitude higher, depending on the coercivity of the sample. For instance in the case of an iron film typical values could be H$_1$= 1 kA/m, H$_0$= 100 A/m, yielding S$_A$/S$_T$= $10^4$. Therefore only in the case of measurements in very high static fields the TOM mode has a sensitivity comparable to the ATOM.

**IV. The ATOM prototype : design criteria and test measurements**

We have built a prototype using a glass beam with an estimated stiffness K=$2.6.10^2$ N/m as the torsion element, Figure 3.reports a scheme of the experimental arrangement We detect the oscillations amplitude at resonance by means of a capacitive sensor integrated with the cantilever having. a 0.15 mm thick air gap.In the present arrangement two main intrinsic sources of synchronous interference signals are present. First, the torque acting on the Helmholtz pair exerted by the static dc magnetic field leads to acustic vibrations. In order to minimize their effect, we have built an excitation coil system with zero net dipole moment and vanishing d$^2$H$_0$/dx$^2$. Vibration insulation is provided through a suitable spring-suspension system and by working in vacuum (or in light gases atmosphere). On the other hand, the coupling between the static magnetic field and ac parasitic magnetic moments arising from eddy currents induced in the capacitor electrodes should also lead to spurious synchronous vibrations. In principle, two possible contributions to ac parasitic magnetic moments should be taken into account, namely: a moment $\tilde{m}_1$ arising from the time-variation of the flux of the H$_y$ excitation field and a moment $\tilde{m}_2$ arising from the time-variation of the flux H$_0$ dc static field due to the oscillation of the cantilever. $\tilde{m}_t$ is by far the main contribution and can be dramatically reduced if the cantilever is oriented parallel with respect to $H_y$.

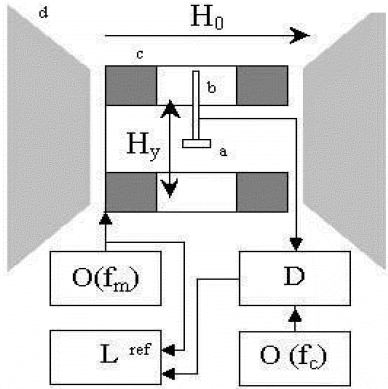

**Fig.3** Basic scheme of the experimental arrangement used: (a) sample, (b) cantilever and sensor, (c) excitation field coil, (d) electromagnet. Electronic components: ($O(f_c)$), carrier frequency oscillator, ($O(f_m)$) modulation frequency oscillator, (D) capacitance radio detector, (L) lock-in amplifier.

As regards the electronic read-out circuit we have developed a capacitance radio-detector of small oscillations [14] which in typical working conditions (Vg=1 V), has an experimentally tested resolution of $\delta Y_L = 0.1$ nm [15, 16]. Substituting this value in eq. (27), one obtains the moment resolution of this prototype, which turns out to be of the order of $10^{-9}$ Am$^2$. This value has been experimentally tested.

In Fig.4 we report the hysteresis loop of a 100 nm thick Fe/Co multilayer measured both with the ATOM prototype and with a commercial extraction magnetometer. The ATOM measurement is by far the most accurate. The high resolution of this measurement allows in the present case to reveal fine details, such as the residual hysteresis.which persists well above the coercive field [17]. In the ATOM measurement we neatly observe a critical field the hysteretic behaviour starts and which we interpret to be due to the onset of a weak stripe domain structure in the multilayer.

### V. The Torsion Induction Magnetometer

The ATOM lends itself to a further application: the *Torsion Induction Magnetometer* (TIM), that comes as a special variant of the operating mode of the instrument. It consists merely in the interchange of the active and passive functions of the excitation and detection systems. In practice one excites the oscillation at resonance of the cantilever by feeding the capacitor and utilizes the coil system as a detector of the induced signal [18]. The device in this configuration is equivalent to a susceptometer for detecting the Reversible Transverse Susceptibility (RTS), in which the exciting field and the induced signals are perpendicular to the bias magnetic field [19, 20]. However

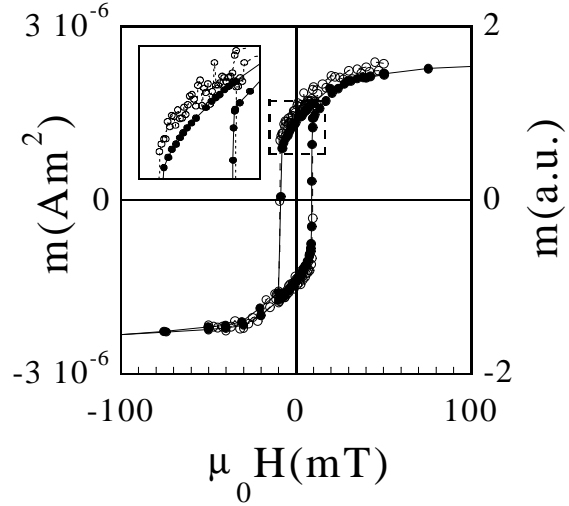

**Fig.4** Hysteresis loop of a 100 nm thick Fe/Co multilayer carried out at room temperature with a commercial extraction magnetometer (open circles and dashed line) and with the ATOM prototype (closed circles and solid line).

there is an important difference that renders the TIM more effective: the fact that the measurement is not performed in the reference frame of the sample but in that of the overall applied magnetic field $H_0$, which actually undergoes oscillatory rotations by a small angle $\varphi=\varphi_0 \cdot e^{i\omega t}$ with respect to the sample. In effect the sample "sees" two magnetic fields: a static $H_0$ field and an oscillating transverse field equal to $H_0\varphi$. In order to understand the effect of such an arrangement let us consider as an example the case of a spherical sample of a random polycristalline uniaxial ferromagnet, such as an isotropic barium ferrite permanent magnet. When the sample performs torsional oscillations, the pick-up coils detect the transverse component of the magnetic moment of the sample, $m_t$, that ought to be zero for $H_0 > H_A$, the anisotropy field, and a finite signal for $H_0 < H_A$. In fact the measured signal is proportional to a transformed RTS, $\chi'_t = (M/H_0) - \chi_t$. It means that the TIM is a kind of transverse susceptometer capable of working in *null mode*. The presence of magnetic stiffness in the sample due to the action of the bias magnetic field causes a shift in the resonance frequency of the torsion pendulum (or cantilever), that could be easily measured [21] Again this magnetic stiffness C vanishes above $H_A$ and provides an independent information on the RTS. In effect the two quantities are connected by the relation $\chi_t = (M/H) - C/(VH^2)$, where V is is the sample volume. In general the TIM is expected to make easier the detection of the singularity at $H=H_A$ and could be advantageously employed to make the most of the great sensitivity of the

RTS to the presence of single domain particles, a peculiarity of particular interest in the permanent magnet and recording materials research area. Actually the SPD theory [20] predicts that the RTS displays a singularity in the form of a cusp at $H=H_A$. However it is known that the real part of RTS does not show any singularity at the anisotropy field in multidomain materials, because the phenomenon is completely screened by the domain wall displacement. The effect is instead present in single domain particles [22]. The proposed methods may be considered in the framework of the extension of the SPD techniques to the study of the singular behavior of the frequency dependent complex susceptibility tensor and subsequent Reversible Higher Order Complex Susceptibilities (RHOCS) [23, 24]. An example is the complex second order Reversible Parallel Susceptibility (RPS) that is measured by detecting the second harmonic of the complex RPS [25]. Finally it has to be noted that even greater sensitivity can be achieved by the TIM by the addition of a modulating coil generating a field at a frequency $\omega_2$, parallel to the bias field. The amplitude modulation of the detected signal is proportional to $d\chi'_t/dH$, which is related with the second order RTS. In this case the above mentioned singularity at $H_A$ is enhanced in the form of a divergence [24, 26]. The application of these SPD techniques to nanostructured materials could in principle reveal the threshold at the nanoscale level where the cooperative effects of the exchange interaction at the grain boundaries give rise to substantial modifications of the anisotropic behaviour of the system.

## VI. Conclusions

We have shown the a resonant torsion oscillation magnetometer utilizing a cantilever beam as a torsion element can achieve high sensitivity if magnetically activated. We expect to achieve significant improvements by carefully studying a microfabricated silicon cantilever and holder displaying much larger compliance, and preferably obtained from one single wafer in order to minimize the mechanical losses due to clamping.

**Acknowledgments**

The authors are indebted to Alberto Bazzini for the help provided in designing and constructing the mechanical parts of the instrument.

**References**

[1] H. Zijlstra, Experimental methods in magnetism, North Holland, Amsterdam, 1967.
[2] C.A. Neugebauer, Phys. Rev., 116 (1958) 1441.
[3] U. Gradmann, W Kummerle, R. Thau, Appl. Phys., 10 (1976) 219.
[4] M.V. Chaparala, O.H. Chung, and M.J. Naughton, AIP Conf. Proc. 273,407, (1993).
[5] C. Rossel, M. Willemin, A. Gasser, H. Bothuizen, G.I. Meijer, and H. Keller Rev. Sci. Instrum.69,3199,(1998).
[6] P.A. Crowell, A. Madouri, M.Specht, G. Chaboussant, D. Mailly, L.P. Levy Rev.Sci. Instrum. 67 ,4161,1996.
[7] Th. Höpfl, D. Sander, H. Höche, J. Kirschner, Rev. Sci. Instrum. 72 (2001) 1495.
[8] M. Lohndorf, J. Moreland, P. Kabos, N. Rizzo, J. Appl. Phys. 87 (2000) 5995.
[9] J. Morillo, Q. Su, B. Panchapakesan, M Wuttig, D. Novotny, Rev. Sci. Instrum. 69 (1998) 3908.
[10] J.G.E. Harris, D.D. Awschalom, F. Matsukura, H.Ohno, K.D. Maranowski, A.C. Gossard, Appl. Phys. Lett. 75 (1999) 1140.
[11] C. Rossel, P. Bauer, D. Zech, J. Hofer, M. Willemin, H. Keller, J.Appl. Phys. 79 (1996) 8166.
[12] R.D. Biggar, J.M. Parpia, Rev.Sci. Instrum. 69 (1998) 3558.
[13] J. Moreland, J. A. Beall, S. E. Roussek, 242-245 (P2) (2002), 1157.
[14] V. Braguinski and A. Manoukine, Mesure de petites forces dans les experiences physiques, MIR, Moscow, 1976.
[15] G. Asti, M. Ghidini, R. Pellicelli, M. Solzi, J. Magn. Magn. Mater. 242-245 (P2) (2002) 984.
[16] G. Asti, M. Ghidini, R. Pellicelli, M. Solzi, unpublished.
[17] G. Asti, M. Carbucicchio, M. Ghidini, M. Rateo, G. Ruggiero, M. Solzi, F. D' Orazio, F. Lucari, J. Appl. Phys. 87 (2000) 6689.
[18] G. Asti, IEEE Trans. Magn. MAG-17 (1981) 2630
[19] A. Aharoni, E. H. Frei, S. Shtrikman and D. Treves, Bull. Res. Council Israel 6A (1957) 215.
[20] G. Asti and S. Rinaldi, J. Appl. Phys. 45 (1974) 3600
[21] C. Voigt and A. Hempel, Phys. Stat. Sol. 33 (1969) 249.
[22] L. Pareti and G. Turilli, J. Appl. Phys. 61 (1987) 5098.
[23] G. Asti, Singular point detection of nanostructured magnets, in: G.C. Hadjipanayis and R.W. Siegel (Eds.), Nanophase Materials: Synthesis, Properties and Applications, Kluwer Publ., Dordrecht, 1994, pag. 691-698.
[24] M. Solzi, M. Ghidini and G. Asti, Macroscopic magnetic properties of nanostructured and nanocomposite systems, in H. S. Nalwa (Ed.), Magnetic Nanostructures, American Scientific Publishers, N.Y., 2002.
[25] G. Asti, M. Ghidini and M. Solzi, IEEE Trans. Magn. MAG-36 (2000) 3605.
[26] R.W. Chantrell, A. Hoare, B. Melville, H.J. Lutke-Stetzkamp and S. Methfessel, IEEE Trans. Magn MAG-25 (1989) 4216.